\documentclass[a4paper,10pt,twoside]{cpc-hepnp}

\usepackage{multicol}
\usepackage{graphicx}
\usepackage{booktabs}
\usepackage{amssymb,bm,mathrsfs,bbm,amscd}
\usepackage[tbtags]{amsmath}
\usepackage{lastpage}
\usepackage{CJK}
\def\half{{\textstyle{1\over 2}}}

\begin{document}
%\begin{CJK*}{GBK}{song}

\fancyhead[c]{\small Submitted to Chinese Physics C} \fancyfoot[C]{\small 010201-\thepage}

\footnotetext[0]{Received July 2015}

\title{$D^*\bar{D}_1(2420)$ and $D\bar{D}'^*(2600)$ interactions and the charged charmonium-like state $Z(4430)$\thanks{Supported by the Major
State Basic Research Development Program in China (No. 2014CB845405),
the National Natural Science Foundation of China (Grants No. 11275235) }}

\author{%
    HE Jun$^{1,3,4}$\email{junhe@impcas.ac.cn} and L\"U Pei-Liang$^{1,2}$}

\maketitle

\address{%
$^1$Theoretical Physics Division, Institute of Modern Physics, Chinese Academy of Sciences,
Lanzhou 730000, China\\
$^2$University of Chinese Academy of Sciences, Beijing 100049, China \\
$^3$Research Center for Hadron and CSR Physics,
Lanzhou University and Institute of Modern Physics of CAS, Lanzhou 730000, China\\
$^4$State Key Laboratory of Theoretical Physics, Institute of
Theoretical Physics, Chinese Academy of Sciences, Beijing  100190,China\\

}

\begin{abstract}
The $D^*\bar{D}_1(2420)$ and $D\bar{D}'^*(2600)$ interactions are
studied in a
one-boson-exchange model.  Isovector bound state solutions with spin parity
$J^P=1^{+}$ are found from the $D^*\bar{D}_1(2420)$ interaction, which
may be related to the observed charged charmonium-like state $Z(4430)$.
There
is no bound state solution found from the $D\bar{D}'^*(2600)$
interaction.
\end{abstract}



\footnotetext[0]{\hspace*{-3mm}\raisebox{0.3ex}{$\scriptstyle\copyright$}2013
Chinese Physical Society and the Institute of High Energy Physics
of the Chinese Academy of Sciences and the Institute
of Modern Physics of the Chinese Academy of Sciences and IOP Publishing Ltd}%

\begin{multicols}{2}

\section{Introduction}

A resonant structure near 4.43 GeV in the $\pi^\pm\psi'$ invariant
mass distribution was first observed by  the Belle
Collaboration~\cite{Choi:2007wga}, and is the first evidence of the
existence of charged charmonium-like states. The mass
$M=4433\pm4{\rm(stat)}\pm2{\rm(syst)}$ MeV and width
$\Gamma=45^{+18}_{-13}{\rm (stat)}^{+30}_{-13}{\rm(syst)}$ MeV were
extracted by using a Breit-Wigner resonance shape. A higher mass
$M=4485^{+22+28}_{-22-11}$ MeV and a larger width
$\Gamma=200^{+41+26}_{-46-35}$ MeV were reported by the Belle
Collaboration through a full amplitude analysis of $B^0\to\psi'
K^+\pi^-$ decay and a spin parity of $J^P=1^+$ was favored over other
hypotheses~\cite{Chilikin:2013tch}. Recently, the LHCb Collaboration
released their new result on the $B^0\to\psi'\pi^- K^+$ decay, which
confirmed the existence of the $1^+$ resonant structure $Z(4430)$  with
a mass $4475\pm7^{+15}_{-25}$ MeV and a width $172\pm13^{+37}_{-34}$
MeV ~\cite{Aaij:2014jqa}.

The $Z(4430)$ was observed in the $\psi'\pi$ invariant mass spectrum,
which suggests that it should be an exotic state beyond the conventional
$c\bar{c}$ picture, which has a neutral charge.  Many theoretical
efforts have been made to understand the internal structure of the
$Z(4430)$ and a number of explanations have been offered. Since the
$Z(4430)$ carries charge, the hybrid interpretation is
excluded~\cite{Branz:2010sh}. It is natural to explain the charge
carrier $Z(4430)$ as  a multiquark system in which, as well as
$c\bar{c}$, there exist other light quarks.  The first type of
multiquark explanation is  the excited
tetraquark~\cite{Li:2007bh,Bracco:2008jj,Guerrieri:2014gfa,Maiani:2014aja,Wang:2014vha,Brodsky:2015wza}
where four quarks are in a color singlet.  Another type of multiquark
explanation is a loosely bound state composed of two charmed
mesons~\cite{Ding:2007ar,Cheung:2007wf}, or charmed
baryons\cite{Qiao:2007ce}. There also exist several nonresonant
explanations, such as the threshold cusp effect~\cite{Rosner:2007mu} and
a cusp in the $D^*\bar{D}_1(2420)$ channel \cite{Bugg:2008wu}.

The $Z(4430)$ mass measured by the Belle
Collaboration ~\cite{Choi:2007wga}, $M=4433\pm4 {\rm(stat)}\pm
2{\rm(syst)}$ MeV, is close to the $D^*\bar{D}_1(2420)$ threshold, so it has been
popular to explain the $Z(4430)$ as a $S$-wave $D^*\bar{D}_1(2420)$ molecular
state with $J^P=0^-$  in the one-boson-exchange (OBE) model~\cite{Close:2010wq,Liu:2008xz}.  A
calculation in the context of the QCD sum rule also
favors the $D^*\bar{D}_1(2420)$  bound state explanation
with spin-parity $0^-$~\cite{Lee:2008tz}. The new Belle and LHCb results suggest the
spin-parity of $Z(4430)$ is $1^+$, however. With such an assignment of spin parity, a new
calculation by Barnes $et\ al.$  suggests that the $Z(4430)$ is either
a $D^*\bar{D}_1$ state dominated by long-range $\pi$ exchange, or a
$D\bar{D}^*(1S, 2S)$ state with short-range
components~\cite{Barnes:2014csa}.  It has also been suggested that the
$Z(4430)$ may be from $S$-wave $D\bar{D}'^*_1(2600)$ interaction
because the $Z(4430)$ mass is very close to the
$D\bar{D}'^*_1(2600)$ threshold~\cite{Ma:2014zua}.

In this paper, the
$D^*\bar{D}_1(2420)$ and $D\bar{D}'^*(2600)$ interactions will be studied by solving the
Bethe-Salpeter equation combined with the one-boson-exchange model. The $Z(4430)$ mass is close to the
threshold of four configurations, $D^*$ $\bar{D}'_1(2430)$, $D^*$$\bar{D}_1(2420)$,
$D$$\bar{D}'^*(2600)$, and ${D}^*$$\bar{D}'^*(2550)$.  The large width of the $D'_1(2430)$, however, $\Gamma=384^{+130}_{-110}$ MeV~\cite{Agashe:2014kda}, which
means a very short lifetime, makes it difficult to bind it and the $D^*$
together to form a bound state with width about $170$ MeV. The
configuration ${D}^*\bar{D}'^*(2550)$ has also been related to the $Z(4430)$ in the literature. However, its threshold is about 100 MeV higher than
the $Z(4430)$ mass.
In this work, the constituents will be treated as stable particles as in the OBE model~\cite{Close:2010wq,Liu:2008xz}. However, the physical widths of $D_1$(2420) and $D'^*$(2600) are about 27 MeV and 93 MeV, respectively. Form factors will be introduced to compensate the self energy effects.  The non-zero width will also introduce the three-body effect, which is not included in the current work considering that the thresholds of the three-body channels, such as $DD\pi$, $D^*D^*\pi$ and $DD^*\pi$, are much lower than the mass of the $Z(4430)$. It is also the reason why the configuration $D^*\bar{D}'^*$ (2500) is excluded.
Since  only loosely bound
states are considered, only two configurations,
$D^*\bar{D}_1(2420)$ and $D\bar{D}'^*_1(2600)$, will be included in the current
calculation.

The paper is organized as follows. In the next section a
theoretical frame will be developed to study the $D^*\bar{D}_1$ and
$D\bar{D}'^*$  interactions (we omit the numbers for the masses, $2420$ and
$2600$ respectively, here and hereafter) by solving the Bethe-Salpeter equation. In Section 3, the potential is derived
with the help of effective Lagrangians from the heavy
quark effective theory. The numerical results are given
in Section 4. A summary is given in the last section.

\section{Bethe-Salpeter equation for vertices}

The Bethe-Salpeter equation is a powerful tool to study bound
state problems such as the deuteron~\cite{VanOrden:1995eg}. A
Bethe-Salpeter formalism was developed and applied to study the
$Y(4274)$ and its decay pattern~\cite{He:2011ed,He:2013oma},
the $\Sigma_c(3250)$ as $D^*_0$ $(2400)N$ system~\cite{He:2012zd} and
the $N(1875)$ as $\Sigma(1385)K$ system~\cite{He:2015yva}. In
Refs.~\cite{He:2014nya,He:2015mja,Ke:2012gm}, the $B\bar{B}^*/D\bar{D}^*$ system
was also studied by solving the Bethe-Salpeter equation with boson
exchange mechanism to explore the possible relationship between the
recently observed $Z_b(10610)$/$Z_c(3900)$ and the
$B\bar{B}^*/D\bar{D}^*$ interaction.  We start from the Bethe-Salpeter equation for vertex
$|\Gamma\rangle$,
\begin{eqnarray}
	|\Gamma\rangle={\cal V} G|\Gamma\rangle,
\end{eqnarray}
where ${\cal V}$ and $G$ are the potential kernel and the
propagator for the two constituents of the system.
The vertex function of the system with two configurations can be written as
\begin{eqnarray}
	|\Gamma\rangle=\Gamma^{D^*\bar{D}_1}|D^*\bar{D}_1\rangle+\Gamma^{D\bar{D}'^*}|D\bar{D}'^*\rangle,
\end{eqnarray}
where $\Gamma_{D^*\bar{D}_1}$ and
$\Gamma_{D\bar{D}'^*}$  are the vertex functions after
separating out the flavor parts $|D^*\bar{D}_1\rangle$ and $|D\bar{D}'^*\rangle$.
In this paper SU(2) symmetry is
considered, so the same vertex function is
used for both configurations.

The explicit flavor structures for isovectors ($T$) or isoscalars ($S$) $|D^*\bar{D}_1\rangle$ are \cite{Liu:2008xz}
\begin{eqnarray}
|D^*\bar{D}_1\rangle^+_T&=&\frac{1}{\sqrt{2}}\big(|D^{*+}\bar{D}^0_1\rangle+c|D^+_1\bar{D}^{*0}\rangle\big),\nonumber\\
|D^*\bar{D}_1\rangle^-_T&=&\frac{1}{\sqrt{2}}\big(|D^{*-}\bar{D}^0_1\rangle+c|D^-_1\bar{D}^{*0}\rangle\big),\nonumber\\
|D^*\bar{D}_1\rangle^0_T&=&\frac{1}{2}\Big[\big(|D^{*+}D^-_1\rangle-|D^{*0}\bar{D}^0_1\rangle\big)\nonumber\\
&+&c\big(|D^+_1D^{*-}\rangle-|D^0_1\bar{D}^{*0}\rangle\big)\Big],\nonumber\\
|D^*\bar{D}_1\rangle^0_S&=&\frac{1}{2}\Big[\big(|D^{*+}D^-_1\rangle+|D^{*0}\bar{D}^0_1\rangle\big)\nonumber\\
&+&c\big(|D^+_1D^{*-}\rangle+|D^0_1\bar{D}^{*0}\rangle\big)\Big],\label{flavor structure}
\end{eqnarray}
\normalsize
where $c=\pm$ corresponds to $C$-parity $C=\mp$. The flavor structure
for $D\bar{D}'^*$ configuration is analogous to that of the
$D^*\bar{D}_1$ configuration.

The vertex function is rewritten as
\begin{eqnarray}
|\Gamma
\rangle=\sum_{i=1}^N\Gamma^{i}\sum_{a=1}^n\delta^{i,a}|i,a\rangle,
\end{eqnarray}
with  $i=1$ or 2 for configuration $D^*D_1$ or $DD'^*$, and $a$ stands for the different
components in a configuration. $\delta^{i,a}$ is the factor for $|i,a\rangle$ in Eq.~(\ref{flavor structure}).
After multiplying $\langle j,b|$ on both sides, the Bethe-Salpeter equation
becomes
\begin{eqnarray}
	\Gamma^{j} =\sum_{i}\tilde{\cal V}^{ji} G^{i}\Gamma^{i}, \ {\rm with}	\ \ \tilde{\cal V}^{ji} =\sum_{b,a}\delta^{j,b}
\delta^{i,a}\langle j,b|V|i,a\rangle.
\end{eqnarray}
The above equation is a coupled-channel equation for the two channels
$D^*\bar{D}_1$ and $D\bar{D}'^*$ involved.

The Bethe-Salpeter equation is a 4-dimensional integral equation. It
is popular to reduce it to a 3-dimensional equation by quasipotential
approximation, and in principle
there exist infinite choices to make the quasipotential approximation.
As in
Ref.~\cite{He:2014nya}, we adopt the covariant spectator
theory~\cite{Gross:1991pm,Gross:2008ps} to make the 3-dimensional reduction.
With the help of the onshellness of
the heavier constituent 2, $D_1/D'^*$, the numerator of the propagator
$P_2^{\mu\nu}=\sum_{\lambda_2}\epsilon_{2\lambda_2}^\mu
\epsilon_{2\lambda_2}^{\nu\dag}$
with $\epsilon^\mu_{2\lambda_2}$ being the polarization vector with
helicity $\lambda_2$. Different from Ref.~\cite{He:2014nya}, where
the off-shell constituent is a pseudoscalar particle $D$, 
constituent 1 here is a vector meson $D^*$. So we will make an approximation
$P_1^{\mu\nu}=\sum_{\lambda_1}\epsilon_{1\lambda_1}^\mu
\epsilon_{1\lambda_1}^{\nu\dag}$ with polarization
$\epsilon^\mu_{1\lambda_1}$ on
shell. Such an approximation will introduce an uncertainty of about several
percent in the numerator of the propagator, which will be further smeared by the
introduction of  form factors which will be given in the next section.
Now, the equation for the vertex is of a form
\begin{eqnarray}
	&&|{\Gamma}^i_{~\lambda_1\lambda_2}\rangle
	=\sum_{j,~\lambda'_1\lambda'_2}{\tilde{\cal{V}}}^{ij}_{\lambda_1\lambda_2\lambda'_1\lambda'_2}
	~{G}^j_0~
	|{\Gamma}^j_{\lambda'_1\lambda'_2}\rangle.
\end{eqnarray}
  Written down in
the center of mass frame where $P=(W,{\bm 0})$, the propagator is
\begin{eqnarray}
	G_0&=&2\pi i\frac{\delta^+(k_2^{~2}-m_2^{2})}{k_1^{~2}-m_1^{2}}
	\nonumber\\&=&2\pi
	i\frac{\delta^+(k^{0}_2-E_2({\bm k}))}{2E_2({\bm k})[(W-E_2({\bm
k}))^2-E_1^{2}({\bm k})]},
\end{eqnarray}
where $k_1=(k_1^{0},\bm
k)=(E_1({\bm k}),\bm k)$, $k_2=(k_2^{0},-\bm
k)=(W-E_1({\bm k}),-\bm k)$ with $E_{1,2}({\bm k})=\sqrt{
m_{1,2}^{~2}+|\bm k|^2}$.

The integral equation  can be written explicitly as
\begin{eqnarray}
&&(W-E^i_1({\bm k})-E^i_2({\bm k}))\phi^i_{\lambda_1\lambda_2}({\bm k})\nonumber\\
&=&\sum_{j,~\lambda'_1\lambda'_2}\int\frac{d{\bm k}'}{(2\pi)^3}
V^{ij}_{\lambda_1\lambda_2\lambda'_1\lambda'_2}({\bm k},{\bm k}',W)
\phi^{j,j}_{\lambda'_1\lambda'_2}({\bm k}'),
\end{eqnarray}
with
\begin{eqnarray}
	&&V^{ij}_{\lambda_1\lambda_2\lambda'_1\lambda_2}({\bm k},{\bm k}',W)\nonumber\\
&=&\frac{i~\bar{\cal
	V}^{ij}_{\lambda_1\lambda_2\lambda'_1\lambda'_2}({\bm k},{\bm
	k}',W)}{\sqrt{2E^i_1({\bm
	k})2E^i_2({\bm k})2E'^j_1({\bm k}')2E'^j_2({\bm k}')}}, \label{Eq:
	Lp}
\end{eqnarray}
where the reduced potential kernel
\begin{eqnarray}
\bar{\cal
	V}^{ij}_{\lambda_1\lambda_2\lambda'_1\lambda'_2}({\bm k},{\bm
	k}',W)=F^i({\bm
	k})\tilde{\cal{V}}^{ij}_{\lambda_1\lambda_2\lambda'_1\lambda'_2}({\bm k},{\bm
	k}',W)F^j({\bm k}'),
\end{eqnarray}
with a factor as
$
F^i({\bm k})=\sqrt{2E^i_2({\bm k})/( W-E^i_1({\bm k})+E^i_2({\bm
k}))}.$
The normalized wave function can be related to the vertex as
$|\phi^i_{\lambda_1\lambda'_2}\rangle=N^i|\psi^i_{\lambda_1\lambda_2}\rangle
=N^i(F^{i})^{-1}G^i_0~|{\Gamma}^i_{\lambda_1\lambda_2}\rangle$ with
the normalization factor
$N^i({\bm k})=\sqrt{2E^i_1({\bm k})E^i_2({\bm k})/(2\pi)^5W}.$

A partial wave expansion
can reduce the 3-dimensional integral equation to a one-dimensional equation,
\begin{eqnarray}
	&&(W-E^k_1(|\bm k|)-E^k_2(|\bm k|))\phi_{k}(|{\bm
	k}|)\nonumber\\&=&\sum_l\int  \frac{|{\bm k}'|^2d|{\bm
	k}'|}{(2\pi)^3}V_{kl}(|{\bm k}'|,|{\bm
	k}'|)\phi_{l}(|{\bm k}'|),\label{Eq: final equation}
\end{eqnarray}
where $k/l$ is the number of wave functions with a certain spin-parity.

\section{Lagrangians and potential}

For a loosely bound system, long-range interaction through the $\pi$
exchange should be more important than  short-range interaction through heavier mesons.  Moreover, in the isovector sector the isospin factors are $-1/2$ and $1/2$ for $\rho$ and $\omega$ mesons, respectively \cite{He:2015mja}. The cancelation between the contributions from these two mesons introduces further suppression of the short-range interaction. Hence, the heavier mesons,
$\rho$ and $\omega$,  are not considered in this paper.
The $\sigma$ exchange which mediates the medium range interaction is included as in Ref.~\cite{Liu:2008xz}. We will find that the $\sigma$ exchange is negligible compared with $\pi$ exchange.

The effective Lagrangians describing the interaction between the
light pseudoscalar meson $\mathbb{P}$ and heavy flavor mesons are constructed with the help of the
chiral symmetry and heavy quark symmetry \cite{Isola:2003fh,Casalbuoni:1996pg},
\begin{eqnarray}\label{eq:lag-p-exch}
\mathcal{L}_{D^*D^*\mathbb{P}}   &=&
  i\frac{g}{f_\pi}\big[-i\epsilon_{\alpha\mu\nu\lambda}D^{*\mu}_b\overleftrightarrow{\partial}^\alpha
  D^{*\lambda\dag}_a\partial^\nu
  \mathbb{P}_{ba}
  \nonumber\\&+&i\epsilon_{\alpha\mu\nu\lambda}\tilde{D}^{*\mu\dag}_a\overleftrightarrow{\partial}^\alpha
  \tilde{D}^{*\lambda}_b\partial^\nu \mathbb{P}_{ab}\big],\\
  \mathcal{L}_{D_1D_1\mathbb{P}}
  &=&i\frac{5k}{6f_\pi}\big[i\epsilon_{\alpha\mu\nu\lambda}D^{\mu}_{1b}
  \overleftrightarrow{\partial}^\alpha
  D^{\lambda\dag}_{1a}\partial^\nu \mathbb{P}_{ba}
  \nonumber\\&-&i\epsilon_{\alpha\mu\nu\lambda}
  \tilde{D}^{\mu\dag}_{1a}\overleftrightarrow{\partial}^\alpha
  \tilde{D}^{\lambda}_{1b}\partial^\nu \mathbb{P}_{ab}\big],\\
  \mathcal{L}_{D^{(')*}D\mathbb{P}} &=&
  \frac{2g^{(')}\sqrt{m_Dm_{ D^{(')*}}}}{f_\pi} \nonumber\\
  &\cdot&\big[-(D_bD^{(')*\dag}_{a\lambda}
  +D^{(')*}_{b\lambda}D^\dag_{a})\partial^\lambda{}\mathbb{P}_{ba}
  \nonumber\\&+&
  (\tilde{D}^{(')*\dag}_{a\lambda}\tilde{D}_b
  +\tilde{D}^\dag_{a}\tilde{D}^{(')*}_{b\lambda})\partial^\lambda\mathbb{P}_{ab}\big],\\
  \mathcal{L}_{D_1D^{(')*}\mathbb{P}}   &=&
  i\sqrt{\frac{2}{3}}\frac{h^{(')}_1+h^{(')}_2}{\Lambda_\chi f_\pi}
  \sqrt{m_{D_1}m_{D^{(')*}}}\nonumber\\&\cdot&
  \Big\{[-\frac{1}{4m_{D_1}m_{D^{(')*}}}D_{1b}^\alpha
  \overleftrightarrow{\partial}^\rho
  \overleftrightarrow{\partial}^\lambda
  D^{(')*\dag}_{\alpha a}\nonumber\\
  &\cdot&\partial_\rho\partial_\lambda \mathbb{P}_{ba}
  -D_{1b}^\alpha  D^{(')*\dag}_{\alpha a}\partial_\rho\partial_\rho
  \mathbb{P}_{ba}\nonumber\\
  &+&3D_{1b}^\alpha
  D^{(')*\dag\beta}_a\partial_\alpha\partial_\beta
  \mathbb{P}_{ba}]\nonumber\\
  &-&[-\frac{1}{4m_{D_1}m_{D{(')*}}}D^{(')*\dag}_{\alpha a}
  \overleftrightarrow{\partial}^\rho
  \overleftrightarrow{\partial}^\lambda D_{1b}^\alpha
  \nonumber\\
  &\cdot&\partial_\rho\partial_\lambda \mathbb{P}_{ab}
-
D^{(')*\dag}_{\alpha
a}D_{1b}^\alpha\partial_\rho\partial_\rho\mathbb{P}_{ab}
\nonumber\\&+&3D^{(')*\dag\beta}_aD_{1b}^\alpha
\partial_\alpha\partial_\beta \mathbb{P}_{ab}]\Big\},
\end{eqnarray}
which corresponds to $D=(D^0,D^+,D_s^+)$
and $\tilde{D}=(\bar{D}^0,D^-,D_s^-)$.
The coupling constant $g$ can be extracted from the experimental $D^*$ width
 with a value $g=0.59$~\cite{Isola:2003fh}.
Falk and Luke obtained an approximate relation
$k=g$ in quark model \cite{Falk:1992cx}. With the available
experimental information, Casalbuoni and coworkers
extracted $h'=(h_1+h_2)/\Lambda_\chi=0.55$
GeV$^{-1}$~\cite{Casalbuoni:1996pg}.
The coupling constant for $D'^*$ decaying into $D\pi$ and $D_1\pi$ can
be extracted from the decay widths obtained in quark model as $\Gamma_{D'^*\to D\pi}=10.84$ MeV
and $\Gamma_{D'^*\to D_1\pi}=0.28$ MeV~\cite{Segovia:2013sxa}.
The values are $g'=0.086$ and  $h''=(h'_1+h'_2)/\Lambda_\chi=0.42$
GeV$^{-1}$. The relative phases between the Lagrangians are not fixed, which will be discussed later.

The $\sigma$ exchange which mediates the medium range interaction is included as in Ref.~\cite{Liu:2008xz}.
The Lagrangians for the scalar $\sigma$ meson read,
\begin{eqnarray}\label{eq:lag-s-exch}
\mathcal{L}_{P^*P^*\sigma}   &=& 2g_\sigma [D^*D^*\sigma+
\tilde{D}^*\tilde{D}^*\sigma],\\
\mathcal{L}_{PP\sigma}   &=&2g_\sigma [-D^*D^*\sigma-
\tilde{D}^*\tilde{D}^*\sigma],\\
\mathcal{L}_{P_1P_1\sigma}
&=&2g'_\sigma[-D_1D_1\sigma-\tilde{D}_1\tilde{D}_1\sigma].
\end{eqnarray}
The coupling constant $g_\sigma=g'_\sigma=-\frac{1}{2\sqrt{6}}g_\pi$ with
$g_\pi=3.73$ \cite{Bardeen:2003kt}.

With the above Lagrangians, we can obtain the potential for direct
and cross diagrams,
\begin{eqnarray}
&&{\cal
V}^{ij}_{\lambda_1\lambda_2,\lambda'_1\lambda'_2}(p_1,p_2;p'_1,p'_2)=I^{ij}_d{\cal
V}^{d~ij}_{\lambda_1\lambda_2,\lambda'_1\lambda'_2}(p_1,p_2;p'_1,p'_2)\nonumber\\&+&I^{ij}_c {\cal
V}^{c~ij}_{\lambda_1,\lambda_2,\lambda'_2\lambda'_1}(p_1,p_2;p'_2,p'_1),
\end{eqnarray}
where $p^{(')}_{1,2}$  is the initial (final) momentum for
constituent 1 or 2.  The flavor factor $I^{ij}_{c,d}$ is listed in
Table~\ref{flavor factor}.
\renewcommand\tabcolsep{0.3cm}
\renewcommand{\arraystretch}{1.}
\begin{center}
\tabcaption{The flavor factors $I^{ij}_d$ and $I^{ij}_c$ for direct and cross diagrams and different exchange mesons.
\label{flavor factor}}

	\begin{tabular}{c|cccc|cc}\bottomrule[2pt]
Isospin &\multicolumn{2}{c}{1}&\multicolumn{2}{c|}{0}&\multicolumn{1}{c}{1}&\multicolumn{1}{c}{0}\\\hline
exchange &$\pi$ &$\sigma$ &$\pi$ &$\sigma$&$\pi$&$\pi$\\\hline
$D^*D_1\to D^*D_1$&$-\frac{1}{2}$&1 &$\frac{3}{2}$&1&$-\frac{1}{2}c$ &$\frac{3}{2}c$ \\
$DD'^*\to DD'^*$&0& 1&0 &1 &$-\frac{1}{2}c$ &$\frac{3}{2}c$\\
$D^*D_1\to DD'^*$&$-\frac{1}{2}$&0 &$\frac{3}{2}$&0&0&0 \\
$DD'^*\to D^*D_1$&$-\half$& 0&$\frac{3}{2}$ &0 & 0&0\\
\toprule[2pt]
\end{tabular}
\end{center}

The form factor is introduced to
compensate the off-shell effect of heavy mesons, and is also required by the convergence of the equation.
It is also convenient to interpret
the form factors as self-energies, which is important in this work due to the large decay width of the heavier constituent, $D_1/D'^*$~\cite{Gross:1991pm}.
In this work, we adopt
\begin{eqnarray}
f(q^2)=[\frac{n\Lambda^4}{n\Lambda^4+(m^2-q^2)^2}]^n. \label{FF1}
\end{eqnarray}
Here $n>2$ is adopted to make the equation convergent. We will present
the results with $n\to\infty$, that is, an exponential type of form factor $f(q^2)\to
e^{-(m^2-q^2)^2/\Lambda^4}$, also to show the sensitivity of results to $n$.
In the propagator of the exchange meson we make a replacement
$q^2\to-|q^2|$ to remove the singularities as in
Ref.~\cite{Gross:2008ps}. The form factor for the light meson is
chosen as a monopole type
$f(q^2)={(\Lambda^2-m^2)}/{(\Lambda^2+|q^2|)}$.  The cut-off can be related to the radius of the hadron $r^2=\frac{6}{f(0)}\frac{d f(q^2)}{dq^2}|_{q^2=0}$, which is about $0.5\sim 1$ fm for a meson. The cut-off is about $1.4\sim2.7$ GeV for exponential type or $0.5\sim1$ GeV for monopole type. Such an estimation is very rough, so in this work we choose the cut-off as a free parameter from 0.8-2 GeV.

\section{Numerical results}

To search for the bound state from the $D^*\bar{D}_1-D\bar{D}'^*$
interactions, the integral equation will
be solved following the procedure in Ref.~\cite{He:2014nya}.  After
discretizing $|{\bm k}|$ and $|{\bm k}'|$  by Gaussian quadrature,
the recursion method in Refs. \cite{Soloveva:2001aa} is adopted to
solve the nonlinear spectral problem. The numerical results are
presented in Fig.~\ref{Fig1}. To show the sensitivity of the results
to parameter $n$ in the form factor in Eq. (\ref{FF1}), the results
with $n=2\to\infty$ are also presented as solid bands. The results
suggest the binding energies are not sensitive to $n$. In this work, all quantum number
$J\leq2$ will be considered in the range of cut-offs $0.8<\Lambda<2$ GeV.

\end{multicols}

\begin{center}
\includegraphics[bb=50 0 960 550,clip, scale=0.48]{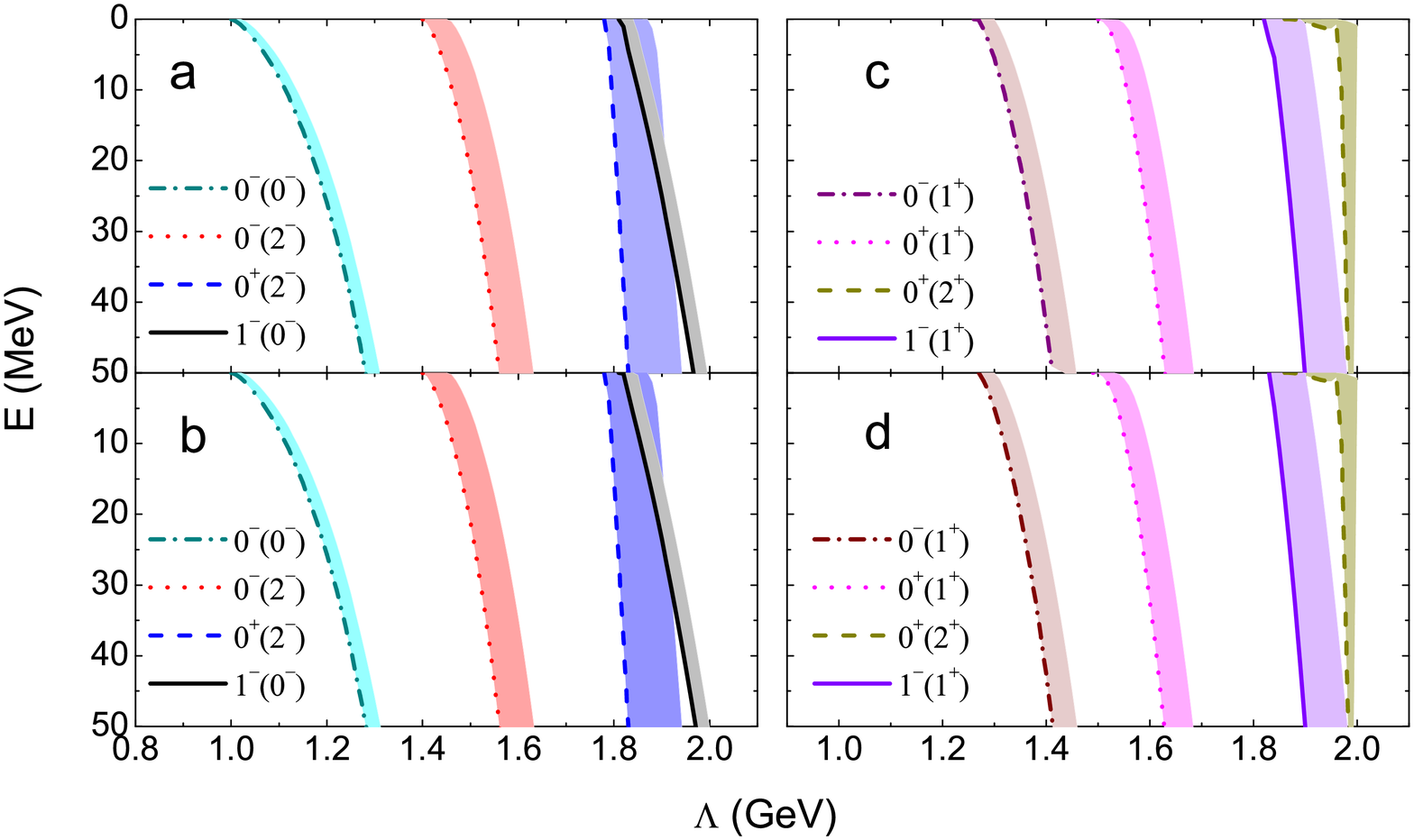}
\figcaption{The binding energies $E$ for the $D^*\bar{D}_1$ system (patterns (a) and (c))
and $D^*\bar{D}_1-D\bar{D}'^*$ system (patterns (b) and (d)) with the variation of
cut-off $\Lambda$. The lines are for the results with
$n=2$ in form factor in Eq. (\ref{FF1}) and the bands  for results with $n=2\to\infty$. \label{Fig1}}
\end{center}

\begin{multicols}{2}

In Fig.~\ref{Fig1}(b) and (d), the coupled-channel results with both
configurations, $D^*\bar{D}_1-D\bar{D}'^*$, are presented, and are
almost the same as these with the configuration $D^*\bar{D}_1$ only (Fig. \ref{Fig1}(a) and Fig. \ref{Fig1}(c)),
which suggests that the $D\bar{D}'^*$ interaction is much weaker than the $D^*\bar{D}_1$ interaction and  transitions between $D^*\bar{D}_1$ and $D\bar{D}'^*$ are negligible.
The $S$ wave $D\bar{D}'^*$ system carries spin-parity $1^+$ which is
consistent with the new experimental results and  the $D\bar{D}'^*$  threshold is very
close to the $Z(4430)$ mass measured in the new LHCb experiment~\cite{Aaij:2014jqa}.
However, in our calculation, no bound state
solution is found from the $D\bar{D}'^*$ interaction with a coupling
constant $h''=0.42$ GeV$^{-1}$.  In this work, the coupling constant $h''$
is determined from the decay width predicted in the quark
model~\cite{Segovia:2013sxa}.   So, we increase the value of $h''^2$ to check if the results are sensitive to $h''^2$ ,  and find that even with $10h''^2$ there is no bound state
produced from  the $D\bar{D}'^*$ interaction.

Different from Ref.~\cite{Liu:2008xz}, the $\pi$ exchange is dominant
in the $D^*\bar{D}_1$ interaction in our model, and the effect of
$\sigma$ exchange is negligible. In the
$\pi$ exchange, the contributions from $D^*\bar{D}_1\to D_1\bar{D}^*$ diagram,
$i.\ e.$, the cross diagram, is much more important than the contribution
from the direct diagram $D^*\bar{D}_1\to D^*\bar{D}_1$.  Hence, the contribution from the
cross diagram  $D^*\bar{D}_1\to D_1\bar{D}^*$  of the $\pi$ exchange is dominant in the coupled $D^*\bar{D}_1-D\bar{D}'^*$
interaction. Since  diagram  $D^*\bar{D}_1\to D_1\bar{D}^*$ is composed of two $D^*D_1\pi$ vertices, the phase of the Lagrangian will be canceled. Hence, its dominance guarantees that the results are not sensitive to the relative phases of the Lagrangians, which are not well fixed.

There exists a bound solution with
quantum number $J^{P}=0^{-}$ with cut-off about 1.8 GeV (see Fig.~\ref{Fig1}(a) and Fig. \ref{Fig1}(b)).
Such an $S$ wave $D^*\bar{D}_1$ molecular state has been related to the
$Z(4430)$ with the assumption that it carries spin parity $J^P=0^-$.
However, the new experimental results favor quantum number $1^+$, which corresponds to
a $P$ wave $D^*\bar{D}_1$ bound state.
In the isovector sector, only two bound states
are produced from the $D^*\bar{D}_1$ interaction. One of them has quantum
number $I^G(J^P)=1^-(1^{+})$ which is consistent with the experimental observed
quantum number of the $Z(4430)$, $J^P=1^+$.

For the coupled $D^*\bar{D}_1-D\bar{D}'^*$ system, the cross diagram contribution from the $\pi$
exchange for channel $D^*\bar{D}_1\to D_1\bar{D}^*$ is dominant.  So the results are only sensitive to the square of coupling
constant, $h'^2$, for $D_1\to D^*\pi$.  The value $h'=0.55$ GeV$^{-1}$ in Ref.~\cite{Casalbuoni:1996pg}
is extracted from the old experimental data, which corresponds to decay
width $\Gamma_{tot}(D_1(2420))\approx 6$ MeV~\cite{Casalbuoni:1996pg}.  Compared with the new
suggested value in the PDG, $25\pm6$ MeV~\cite{Agashe:2014kda}, the
largest possible value of $h'$ is about 1 GeV$^{-1}$. It is of
interest to check the variation of results, especially for bound states
with the $Z(4430)$ quantum numbers, with the variation of coupling constant
$h'$. The results are presented in   Fig.~\ref{Fig2}.

\begin{center}
\includegraphics[bb=60 0 990 700,clip, scale=0.24]{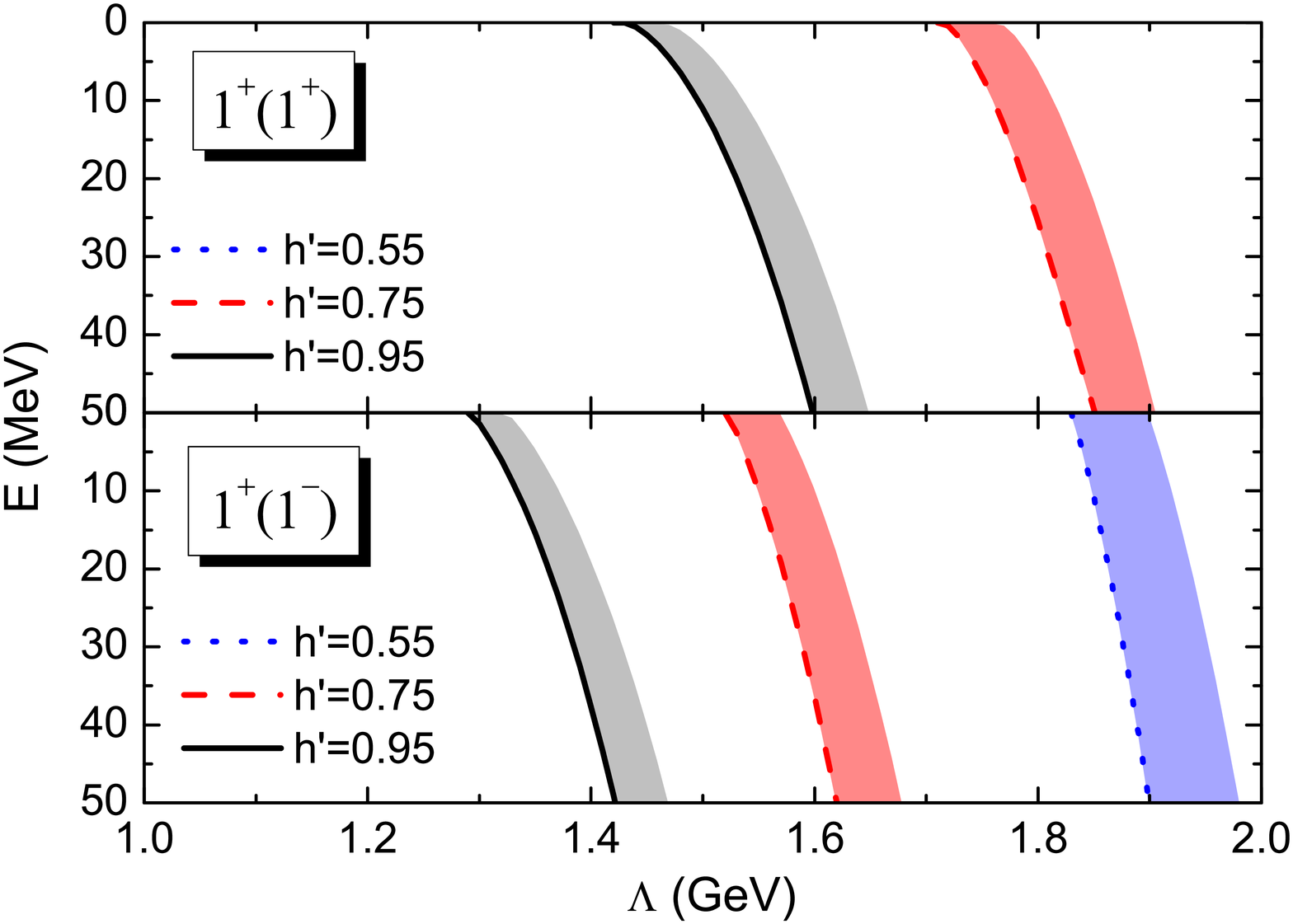}
\figcaption{The binding energies $E$ for  coupled $D^*\bar{D}_1-D\bar{D}'^*$ system with the variation of
cut-off $\Lambda$. The lines are for the results with
$n=2$ in form factor in Eq. (\ref{FF1}) and the bands  for results with $n=2\to\infty$.\label{Fig2}}
\end{center}

With larger $h'$, bound states are generated from the $D^*\bar{D}_1-D\bar{D}'^*$ interactions with smaller cut-offs. For example, with a coupling
constant $h'=0.95$, the isovector bound states with $J^{PC}=1^{++}$ and
$J^{PC}=1^{+-}$ are generated with cut-offs about 1.3 GeV and 1.5 GeV,
respectively.

\section{Summary}

The new experimental results released by the LHCb Collaboration exclude the
$S$ wave $D^*\bar{D}_1$ molecular state interpretation  with quantum
number $J^P=0^-$ for the $Z(4430)$. In this paper we discuss the
possibility to interpret the $Z_c(4430)$ as $D^*\bar{D}_1$ or
$D\bar{D}'^*$ molecular state with quantum number $J^P=1^+$.

Isovector bound state solutions with spin-parity
$J^P=1^{+}$ are found from the $D^*\bar{D}_1(2420)$ interaction, which
may be related to the charged charmonium-like state $Z(4430)$. Different from the Belle
experiment~\cite{Choi:2007wga}, the new observed mass of $Z(4430)$ is above the $D^*\bar{D}_1(2420)$  threshold. However, considering the current large  uncertainties and broad width, further more precise measurements are expected. The $Z(4430)$ is still a candidate for the $D^*\bar{D}_1(2420)$ molecular state. On the theoretical side, it is interesting  to consider the possibility of interpreting the $Z(4430)$ as a resonance from the $D^*\bar{D}_1(2420)$ interaction \cite{He:2015mja}, which can provide a mass above the threshold and is still consistent with the conclusion in this work.

There
is no bound state solution found from the $D\bar{D}'^*(2600)$
interaction.
A calculation with the coupled
$D^*\bar{D}_1-D\bar{D}'^*$ interaction is also
performed, and it is found that the results are almost the same as those obtained from the
$D^*\bar{D}_1$ configuration only.

The current work is performed with the assumption that only channels with thresholds close to the mass of the $Z(4430)$ are important. A more comprehensive study with more coupled channels and more sophisticated treatment of the non-zero width of the excited $D$ meson  will be helpful to further understand the internal structure of the $Z(4430)$ and the molecular states with two excited $D$ mesons.
 
In this work many molecular states are found from the $D^*\bar{D}_1(2420)$ interaction, but only one of them can be related to the observed $Z(4430)$.
This is not surprising because those states are not obtained with the same cut-off, which should be the same for a given interaction. Hence, the states obtained in this work 
do not exist simultaneously. Besides, the effect of some molecular state predicted states may be too small to be observed in current experiments. Further more precise experiments are expected to check their existence.

\ \\

\end{multicols}

\clearpage

%\end{CJK*}
\end{document}